\documentclass[reprint,amsmath,amssymb,aps]{revtex4-1}
\usepackage{graphicx}
\usepackage{color}
\usepackage{amsmath}
\usepackage{amssymb}
\usepackage{amsthm}
\usepackage{booktabs}
\usepackage{float}
\usepackage[dvipdfm,colorlinks,urlcolor=blue,linkcolor=blue,anchorcolor=blue,citecolor=blue]{hyperref}
\usepackage{lineno}

\begin{document}

\title{Multi-particle-collision simulation of heat transfer in low-dimensional fluids}

\author{Rongxiang Luo$^{1,2}$}
\email{phyluorx@fzu.edu.cn}
\author{Stefano Lepri$^{3,4}$}
\email{stefano.lepri@isc.cnr.it}

\newcommand{\fuzhou}{Department of Physics, Fuzhou University, Fuzhou 350108, Fujian, China}
\newcommand{\fujian}{Fujian Science and Technology Innovation Laboratory for Optoelectronic Information of China, Fuzhou 350108, Fujian, China}
\newcommand{\Italy}{Consiglio Nazionale delle Ricerche, Istituto dei Sistemi Complessi, via Madonna del Piano 10, I-50019 Sesto Fiorentino, Italy}
\newcommand{\italy}{Istituto Nazionale di Fisica Nucleare, Sezione di Firenze, via G. Sansone 1, I-50019 Sesto Fiorentino, Italy}

\affiliation{
$^1$ Department of Physics, Fuzhou University, Fuzhou 350108, Fujian, China\\
$^2$ Fujian Science and Technology Innovation Laboratory for Optoelectronic Information of China, Fuzhou 350108, Fujian, China\\
$^3$ Consiglio Nazionale delle Ricerche, Istituto dei Sistemi Complessi, via Madonna del Piano 10, I-50019 Sesto Fiorentino, Italy\\
$^4$ Istituto Nazionale di Fisica Nucleare, Sezione di Firenze, via G. Sansone 1, I-50019 Sesto Fiorentino, Italy
}

\date{\today }

\begin{abstract}
Simulation of transport properties of confined, low-dimensional fluids can be performed efficiently by
means of Multi-Particle Collision (MPC) dynamics with suitable thermal-wall boundary conditions.
We illustrate the effectiveness of the method by studying dimensionality effects
and size-dependence of thermal conduction, properties of crucial importance for understanding
heat transfer at the micro-nanoscale.
We provide a sound numerical evidence that the simple MPC fluid
displays the features previously predicted from hydrodynamics of lattice systems: (1) in 1D, the thermal conductivity $\kappa$ diverges with the system size $L$ as $\kappa\sim L^{1/3}$ and its total heat current autocorrelation function $C(t)$ decays with the time $t$ as $C(t)\sim t^{-2/3}$; (2) in 2D, $\kappa$ diverges with $L$ as $\kappa\sim \mathrm{ln} (L)$ and its $C(t)$ decays with $t$ as $C(t)\sim t^{-1}$; (3) in 3D, its $\kappa$ is independent with $L$ and its $C(t)$ decays with $t$ as $C(t)\sim t^{-3/2}$.
For weak interaction (the nearly integrable case) in 1D and 2D, there exists an intermediate regime of
sizes where kinetic effects dominate and transport is diffusive before crossing over
to expected anomalous regime. The crossover can be studied by decomposing the heat current
in two contributions, which allows for a very accurate test of the predictions.
In addition, we also show that upon increasing the aspect ratio of the system, there exists a dimensional crossover from 2D or 3D dimensional behavior to the
1D one. Finally, we show that an applied magnetic field
renders the transport normal, indicating that pseudomomentum conservation is not sufficient
for the anomalous heat conduction behavior to occur.
\end{abstract}

\maketitle

\section{Introduction}

Simulation is often the only viable tool to study many-particle
systems driven away from equilibrium, especially where
external mechanical and thermal forces are strong.
Molecular dynamics is the most natural approach, but may
be computationally expensive. Since one is often interested
in large-scale properties,  details of microscopic interactions
may not be essential,  since only the basic conservation laws should
matter.  It is thus sensible  to look
for methods based on stochastic processes that
may effectively account for molecular interactions.
One popular approach is the Multi-Particle Collision (MPC) dynamics,
a mesoscale description where individual particles undergo stochastic collisions,
rather than genuine Newtonian forces. The implementation was originally proposed by Malevanets and
Kapral \cite{1999Malevanets,kapral08} and consists of two distinct
stages: a free streaming and a collision one.
Collisions occur at fixed discrete time intervals,
and space is discretized into cells that
define the collision range.
The method captures both thermal fluctuations
and hydrodynamic interactions.

The MPC dynamics is a useful
tool to investigate concrete systems and indeed has been
used in the simulation of a variety of problems, like polymers in solution \cite{1999Malevanets},
colloidal fluids \cite{2009Multi}, plasmas  and even dense stellar systems \cite{di2021introducing} etc.  Besides its computational
convenience it is  also a useful approach to
address fundamental problems in statistical physics
\cite{2011Foffi,benenti2014thermoelectric,luo2020onsager,2022luo}  and
in particular the effect of external sources.

In this work, we focus on the application of the MPC method to study heat transfer in a simple fluid both at equilibrium and in the presence of external reservoirs. In particular, we analyze the dimensionality effects on thermal transport in mesoscopic and confined fluids. Energy transport in low-dimensional systems has been thoroughly studied in recent decades and is crucial for achieving an understanding of macroscopic irreversible heat transfer on the nanoscale \cite{2003LEPRI,2008Dhar,2016Lepri,benenti2020anomalous}. Also, it serves as a theoretical foundation for thermal energy control and management \cite{Li2012,2018Gu,maldovan2013}. This is even more relevant at the nano- and microscale, where novel effects caused by reduced dimensionality, disorder, and nano-structuring affect natural and artificial materials \cite{benenti2023non}. One remarkable property is that in low-dimensional many-particle systems, energy propagates super-diffusively, implying a breakdown of classical Fourier's law. This has been well studied, mostly in lattice systems but much less in fluids. We will show that the MPC dynamics is particularly effective and allows for a very accurate test of existing theories of anomalous transport.

The paper is organized as follows. In Section \ref{sec:mpc} we recall the
basic definitions of the MPC dynamics and the thermal-wall method
used to enforce the interaction with external heat baths.
We then present results of both equilibrium and nonequilibrium for the one- (1D), two- (2D), and three-dimensional (3D) mesoscopic fluids, Sections \ref{sec:1D} to \ref{sec:3D} respectively.
In Section \ref{sec:cross} we illustrate how, upon changing the aspect ratio of the simulation box, one can observe crossover behavior in size dependence of the transport coefficient.
In Section \ref{sec:magnetic}, we briefly
discuss the case 2D fluids with magnetic field to clarify that pseudomomentum conservation is not the necessary condition for the anomalous heat conduction behavior.

\section{Mesoscopic fluid and thermal walls}
\label{sec:mpc}

\begin{figure}[t]
\centering
\includegraphics[width=7cm]{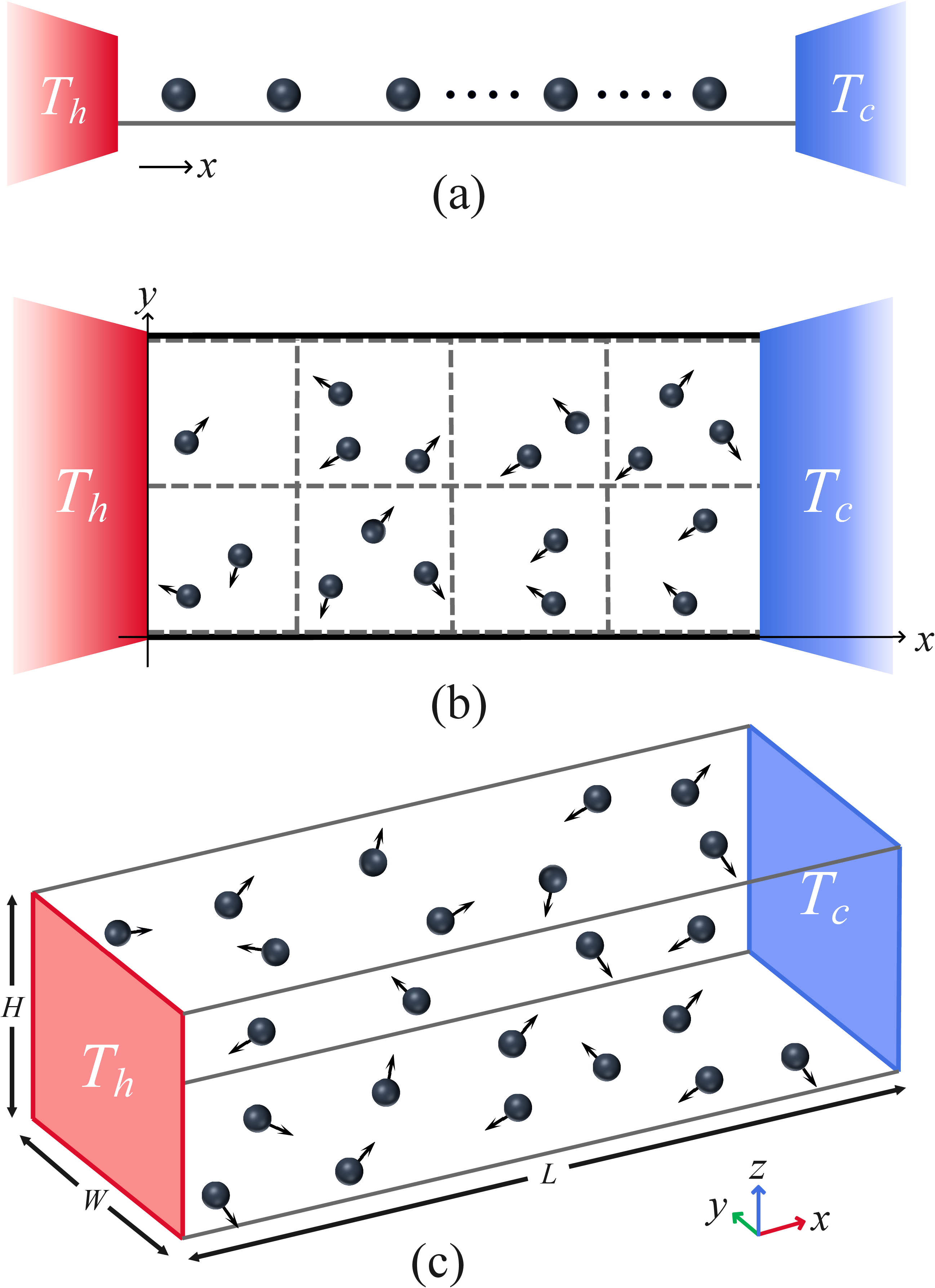}
\caption{(Color online) Schematic illustration of 1D (a) , 2D (b) and 3D (c) fluid of interacting particles in a system volume described by the MPC dynamics. The system is coupled at its left and right ends to one of two heat baths at fixed temperature $T_h$ and $T_c$ (See text for more details).}
\label{model}
\end{figure}

The 1D, 2D, and 3D mesoscopic fluid models we cosider in this study are shown in Fig.~\ref{model}. The fluid consists of $N$ interacting point particles with equal mass $m$, confined in a system volume. As is shown: In 1D system, its volume has a length of $L$  in the $x$ direction; in 2D system, its volume also has a width of $W$ in the $y$ direction; in 3D system, its volume has an additional height of $H$ in the $z$ direction.

At the boundaries $x=0$ and $x=L$, the particles interact with a heat bath of temperature $T_h$ or $T_c$; the heat baths are modeled as thermal-walls~\cite{1978Transport,1998Tehver}. When a particle crosses the $x=0$ (or $x=L$) boundary, it is reflected back with a new random velocity ($v_x$, $v_y$, and $v_z$ in the $x$, $y$, and $z$ directions) assigned by sampling from a given distribution~\cite{1978Transport,1998Tehver}:
\begin{equation}\label{Eqfv}
\begin{aligned}
  f\left(v_{x}\right)= & \frac{ m|v_{x}|}{k_{B}T_\alpha}\textrm{exp}\left(-\frac{mv^{2}_{x}}{2k_{B}T_\alpha}\right), \\
 f(v_{y,z})= &\sqrt{\frac{m}{2\pi k_{B}T_\alpha}}\textrm{exp}\left(-\frac{mv^{2}_{y,z}}{2k_{B}T_\alpha}\right),
\end{aligned}
\end{equation}
where $T_\alpha$ ($\alpha=h, c$) is the temperature of the heat bath in dimensionless units and $k_{B}$ is the Boltzmann constant. The particles are subject to periodic boundary conditions in the $y$ and $z$ directions. We point out that  the numerical results also apply to fixed boundary conditions since, in both cases, $v_{y,z}>0$ and $v_{y,z}<0$ with equal probability $p=0.5$.\par

Interaction among particles as prescribed by the  MPC method~\cite{1999Malevanets,2009Multi,2006Padding}, amounts to first partition the
simulation box  into many cells of linear size
$a$ (as is shown in Fig.~\ref{model} (b) for 2D case). The dynamics evolves
in discrete time steps, each step consisting of a free propagation during a time interval $\tau$ followed by an instantaneous collision event. During propagation, the velocity  $\textbf{\emph{v}}_i$ of a particle is unchanged, and its position is updated as
\begin{equation}\label{Eqr}
\textbf{\emph{r}}_i\rightarrow\textbf{\emph{r}}_i+\tau\textbf{\emph{v}}_i.
\end{equation}
For all particles in a given
cell, their velocities are updated according to the following collision rules:
\begin{itemize}
  \item In the 1D case, the velocity of the $i$th particle in the $j$th cell is changed according to the update rule
   \begin{equation}\label{EqMPC1}
   \textbf{\emph{v}}_i\rightarrow A_j\omega_i+B_j,
   \end{equation}
   where $\omega_i$ is randomly sampled by a thermal distribution at the cell kinetic temperature  $T_j$, while $A_j$ and $B_j$ are cell-dependent parameters, determined by the condition of total momentum and total energy conservation in the cell~\cite{2015Cintio}.
  \item In the 2D case, all particles found in the same cell are rotated around the $z$ axis, with respect to their center of mass velocity $\textbf{\emph{V}}_{\textrm{c.m.}}$ by two angles, $\theta$ or $-\theta$, randomly chosen with equal probability. The velocity of the $i$th particle in a cell is thus updated as
  \begin{equation}\label{EqMPC2} \textbf{\emph{v}}_i\rightarrow\textbf{\emph{V}}_{\textrm{c.m.}}+\hat{\mathcal{R}}^{\pm\theta}\left(\textbf{\emph{v}}_i-\textbf{\emph{V}}_{\textrm{c.m.}}\right),
  \end{equation}
where $\hat{\mathcal{R}}^{\pm\theta}$ are the rotation operators by the angle $\pm\theta$.
  \item In the 3D case, the velocity of the $i$th particle in a cell is updated as in 2D case,
  with the  difference that the rotation axis is also randomly selected.
\end{itemize}
The resulting motion conserves the total momentum and energy of the fluid. Note that the angle $\theta=\pi/2$ for 2D and 3D cases corresponds to the most efficient mixing of the particle momenta. Note also that the probability of collision between particles increases as $\tau$ decreases.
Thus the time interval $\tau$ can be changed to tune the strength of the interactions that,
in turn, will affect the transport properties. \par

In the nonequilibrium setup, we set $T_\alpha$ to be slightly biased from the nominal temperature $T$, i.e., $T_{h,c}=T\pm\Delta T/2$, to investigate the dependence of the thermal conductivity $\kappa$ on the system length $L$.
 In our simulations, each particle is initially given by a random position uniformly distribution and a random velocity generated from the Maxwellian distribution at the temperature $T$. After the system reaches the steady state, we compute the thermal current $J$ that crosses the system according to its definition (i.e., the average energy exchanged in the unit time and unit area between particles and heat bath) and taking it into Fourier's law, $\kappa=JL/(T_h-T_c)$, to compute $\kappa$.
We thus examine the dependence of $\kappa$ on $L$, to assess
whether the heat conduction behavior of the system is anomalous or normal.

In the integrable case (i.e., when $\tau=\infty$, each particle maintains unchanged velocity as it crosses the system from one heat bath to the other), we can obtain an analytical expression for the thermal conductivity:
\begin{equation}\label{EqKL}
\kappa= L\left(d+1\right)\sqrt{\frac{\rho^2k^{3}_B}{2\pi m }}/\left(\frac{1}{\sqrt{T_h}}+\frac{1}{\sqrt{T_c}}\right).
\end{equation}
where  $d$ is the spatial dimension and $\rho$ is the particle density. It is shown that in the integrable case, transport is ballistic and the thermal conductivity is a linear function of the system length. This analytical result will be used to compare with our simulations as a numerical verification.

In the nonequilibrium setting, the difference between normal and abnormal heat conduction behaviors can be also appreciated upon examining the steady-state kinetic temperature profiles $T(x)$. In our simulations, $T(x)$ is measured as described in~\cite{2020Luo}. For systems with normal heat conduction, $T(x)$ is determined by solving the stationary heat equation assuming that the thermal conductivity is proportional to $\sqrt{T}$ as prescribed by standard kinetic theory, yielding~\cite{2001Dhar}
\begin{equation}\label{EqTx}
T(x)=\left[T_{h}^{3/2}\left(1-\frac{x}{L}\right)+T_{c}^{3/2}\frac{x}{L}\right]^{2/3}.
\end{equation}
this prediction will be used to compare with our simulation results for normal heat conduction. On the other hand, for systems with abnormal heat conduction, $T(x)$ is expected to be qualitatively different, being solution of a fractional diffusion equation as demonstrated in several examples ~\cite{2011Lepri,2019Kundu}. A typical feature is that the temperature profile is concave upwards in part of the system and concave downwards elsewhere, and this is true even for small temperature differences~\cite{2000Lippi,2007Mai,2009Lepri}. This will be used also to check our numerical simulations for abnormal heat conduction.\par

To check the results obtained in the nonequilibrium modeling, we will further turn to the comparison with linear-response results obtained in equilibrium modeling. Based on the celebrated Green-Kubo formula, which relates transport coefficients to the current time-correlation functions $C(t)$, the thermal conductivity can be expressed as~\cite{1991Kubo,2003LEPRI,2008Dhar}
\begin{equation}\label{EqGKKL}
\kappa_{\mathrm{GK}}=\frac{\rho}{k_BT^2}\lim_{\tau_{\mathrm{tr}}\rightarrow\infty}\lim_{N\rightarrow\infty}\frac{1}{N}\int_{0}^{\tau_{\mathrm{tr}}}C(t)dt.
\end{equation}
In this formula, $C(t)\equiv\left<\mathcal{J}(0)\mathcal{J}(t)\right>$ and $\mathcal{J}\equiv\frac{1}{2} \Sigma_{i}^{N}\mathbf{v}_{i}^2v_{x,i}$ represents the total heat current along the $x$ coordinate in the equilibrium state. In the simulations, we consider an isolated fluid with periodic boundary conditions also in the $x$ direction. The initial condition is randomly assigned with the constraints that the total momentum is zero and the total energy corresponds to $T$. The system is then evolved and after the equilibrium state is attained, we compute $C(t)$ and the integral in~Eq.~(\ref{EqGKKL}). Usually, the integral is  truncated up to $\tau_{\mathrm{tr}}=L/v_{s}$ ($v_{s}$ is the sound speed)~\cite{2003LEPRI,2008Dhar}. This results the superdiffusive heat transport $\kappa_{\mathrm{GK}}\sim L^{1-\lambda}$ as long as $C(t)$ decays as $\sim t^{-\lambda}$ with $\lambda <1$.\par

In order to compare $\kappa$ and $\kappa_{\mathrm{GK}}$ more accurately, we can resort to the spatiotemporal correlation function of local heat currents to compute the sound speed $v_{s}$ of the system in Eq.~(\ref{EqGKKL}). The spatiotemporal correlation function of local heat currents is defined as~\cite{2003Casati,2006Zhao,2011Levashov}
\begin{equation}\label{EqCxt}
C\left(x,t\right)\equiv \left<\mathcal{J}^{loc}(0,0)\mathcal{J}^{loc}(x,t)\right>.
\end{equation}
Numerically, we compute $C\left(x,t\right)$ as performed in~\cite{2014Chen}: the system is divided into $\frac{L}{b}$ bins in space of equal width $b=0.2$; the local heat current in the $k$\textmd{th} bin and at time $t$ is defined as $\mathcal{J}^{loc}(x,t)\equiv\sum_{i}\frac{1}{2}m_i\textbf{\emph{v}}^{2}_iv_{x,i}$, where $x\equiv kb$ and the summation is taken over all particles that reside in the $k$\textmd{th} bin.
It is  found that $C\left(x,t\right)$ features a pair of pulses moving oppositely away from $x=0$ at the sound speed~\cite{1998Lepri,2005Prosen}, which are recognized to be the hydrodynamic sound modes. Their moving speeds of the two pulses
allows to estimate the sound speed $v_{s}$ within the fluid. \par

As for the choice of parameters, in both  nonequilibrium and equilibrium settings,
we set $T=1$, $\Delta T=0.2$, $m=k_B=a=1$, $\theta=\pi/2$, and $\rho=5$ throughout
the paper. In addition, for all data points shown in the figures, the errors are $\leq 1\%$.\par

\section{1D fluid}
\label{sec:1D}
\subsection{Nonequilibrium results}
\begin{figure}
\centering
\includegraphics[width=9cm]{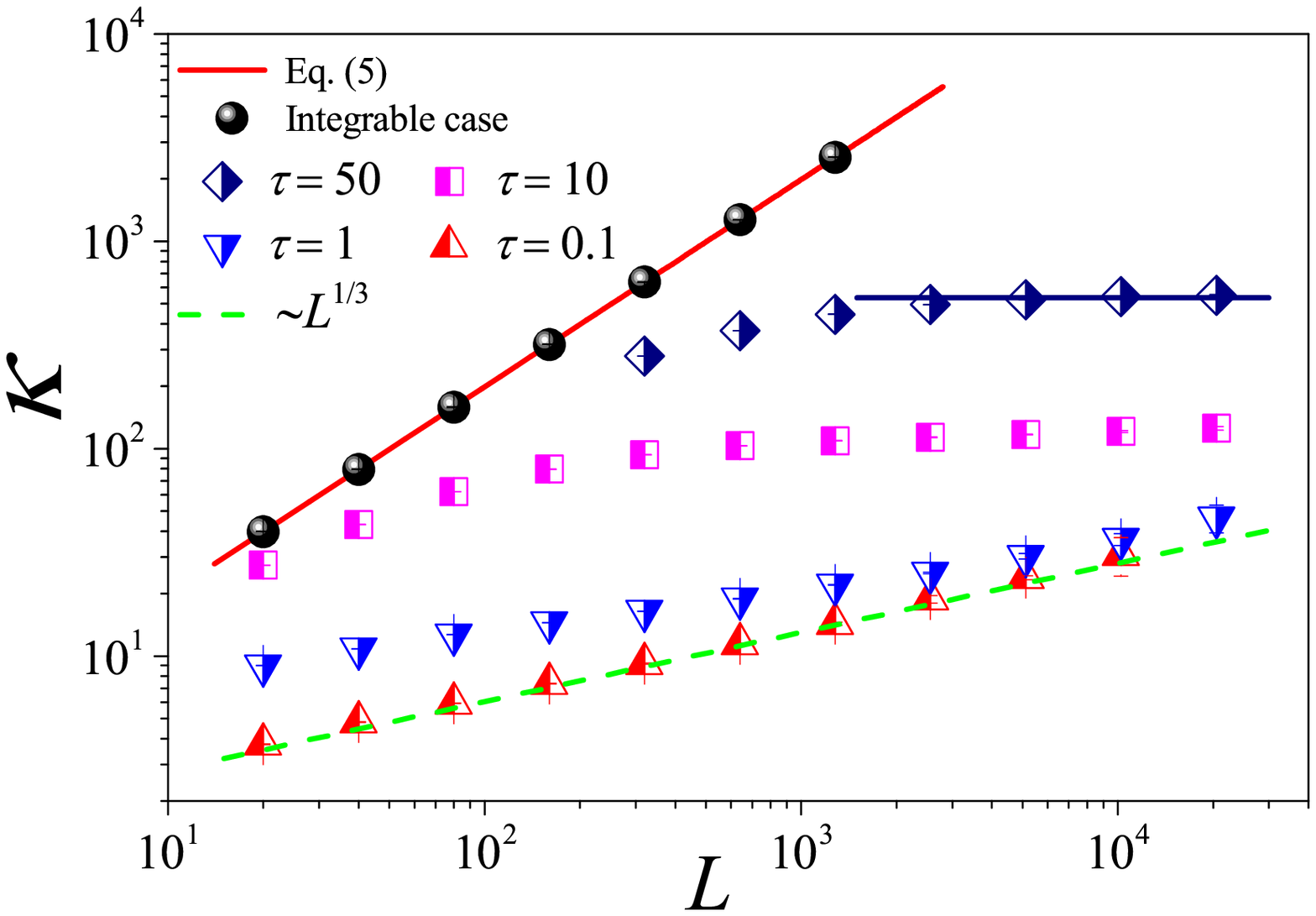}[t]
\caption{(Color online) The thermal conductivity $\kappa$ as a function of the system length $L$ for the 1D fluid system with different $\tau$ values. The symbols are for the numerical results, and for reference the green dashed line indicates the divergence with $L$ as $\sim L^{1/3}$. For $\tau=50$, the horizontal line denotes the saturation value of $\kappa_{\mathrm{GK}}$ obtained by Eq.~(\ref{EqGKKL}), where the integration is up a time $L/v_s$ with $v_s$ measured in Fig.~\ref{figCxt1}.}
\label{figkL1}
\end{figure}

To start, let us discuss the dependence of the thermal conductivity $\kappa$ on the system length $L$ with different interaction strengths. Here, the time interval $\tau$ between successive collisions will be used to tune the strength of the interactions.  The particle mean free path $\ell$,  in an uniform system,
is proportional to the thermal velocity of the fluid $v_T = \sqrt{T}$
and the time of the MPC move, namely
$\ell \sim v_T \tau  $.


To check the results and provide a numerical example,
we first quantify the noninteracting system $\tau= \infty$ . In Fig.~\ref{figkL1}, we report $\kappa$ of 1D case (Eq.~(\ref{EqKL})) with a red line is compared with our simulations (black cycles). It can be seen that they agree very well with each other. These simulations clearly strongly support our analysis.\par

We next turn to the interacting systems with $\tau < \infty$: It can be seen in Fig.~\ref{figkL1} that for  weak interactions ($\tau=50, 10$), $\kappa$ tends to saturate and becomes constant as $L$ is increased, following the Fourier law. However, it can also be seen that for the strong interactions ($\tau=1, 0.1$), $\kappa$ is no longer constant but diverges with $L$. In particular, for $\tau=0.1$, $\kappa$ eventually approaches the scaling $\kappa\sim L^{1/3}$  like that predicted in 1D momentum-conserving fluids~\cite{2002Narayan}. This is the anomalous heat conduction behavior dominated by hydrodynamic effect, well known in nonequilibrium heat transport.

Altogether this can me understood as a crossover from ballistic,
to diffusive (kinetic) and then to anomalous behavior
controlled by corresponding timescales, see
~\cite{2014chen1,2018Zhao,2019Miron,2020Lepri,2021Zhao,2021Lepri,fu2023nonintegrability}.

\begin{figure}
\includegraphics[width=9cm]{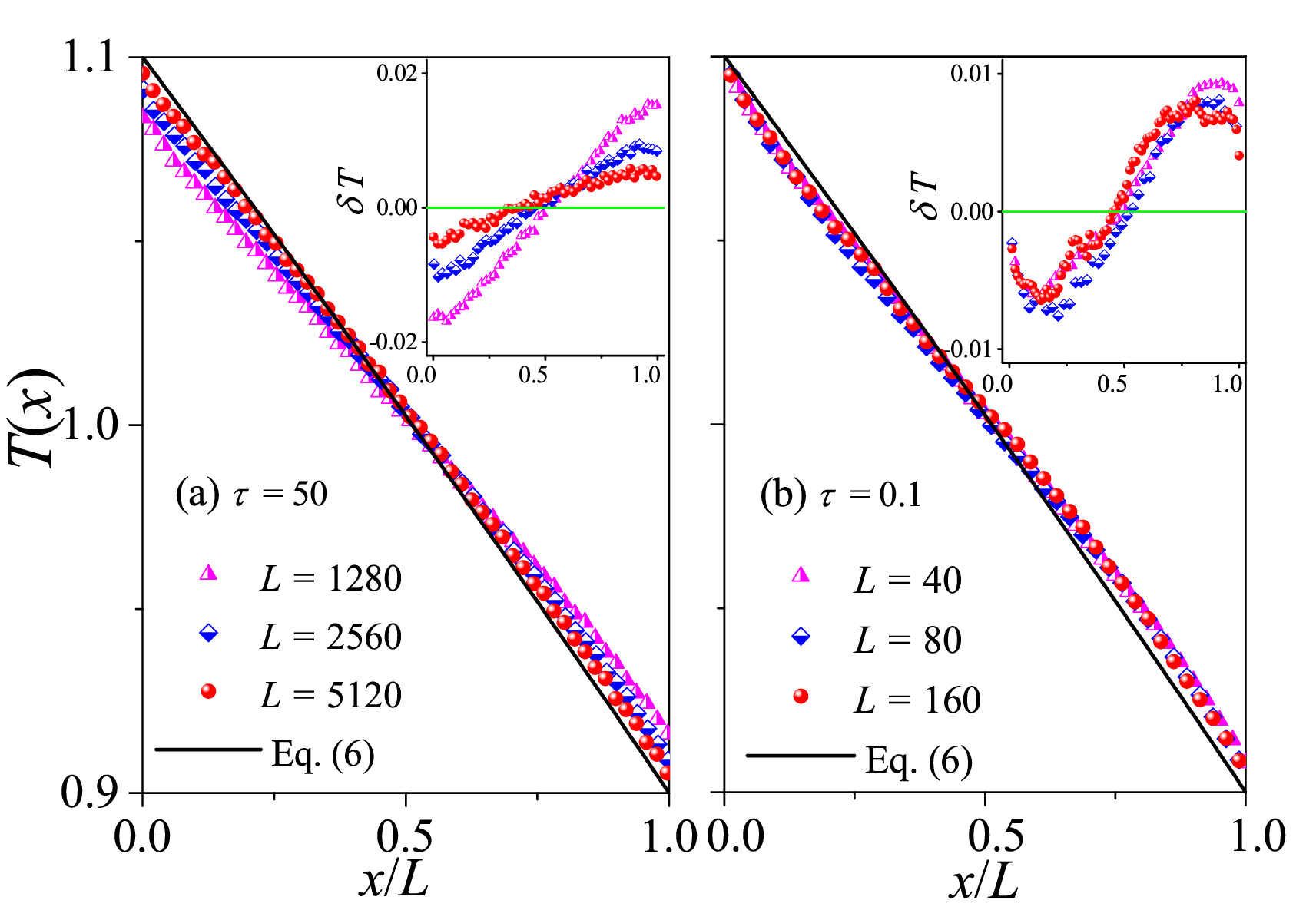}
\caption{(Color online) Plot of temperature profiles $T(x)$ for the 1D fluid system with different $L$ values. Here our numerical results are compared with the analytical Eq.~(\ref{EqTx}). In (a) and (b) we fix $\tau=50$ and $\tau=0.1$, respectively. Inset: Plot of the differences $\delta T$ between the data and the black line, and the green line at $\delta T = 0$ are for reference.}
\label{figTx1}
\end{figure}

The difference between normal and abnormal heat conduction behaviors can be further appreciated also in the steady-state kinetic temperature profiles $T(x)$. For systems with normal heat conduction, $T(x)$ is predicted by Eq.~(\ref{EqTx}). In Fig.~\ref{figTx1}(a) this prediction is compared with our simulation results for $\tau=50$. It is seen that there is a good agreement between the results of our numerical simulations and Eq.~(\ref{EqTx}). To better appreciate the deviations from the prediction, we also plot the differences $\delta T$ between the data and the black line. It is shown in the inset of Fig.~\ref{figTx1}(a) that $|\delta T|$ decreases with increasing $L$, as expected since $|\nabla T|=\Delta T/L$ decreases when $L$ increases, indicating that the linear response can correctly describe the transport properties of the system for large enough system length. As introduced in the section \ref{sec:mpc}, for systems with abnormal heat conduction, the typical feature of the temperature profile is that $T(x)$ is concave upwards in part of the system and concave downwards elsewhere, and this is true even for small temperature differences~\cite{2000Lippi,2007Mai,2009Lepri}. This is confirmed in Fig.~\ref{figTx1}(b) by our numerical simulations for $\tau=0.1$. Note that the data for three different $L$ overlap with each other, implying that the deviations from Fourier's behavior are not finite-size effects. Altogether, those numerical results again support our findings based on the length-dependence of the thermal conductivity, that heat conduction in the weak interactions is normal while in the stronger interactions it is abnormal.\par

\subsection{Equilibrium results}

\begin{figure}
\includegraphics[width=9cm]{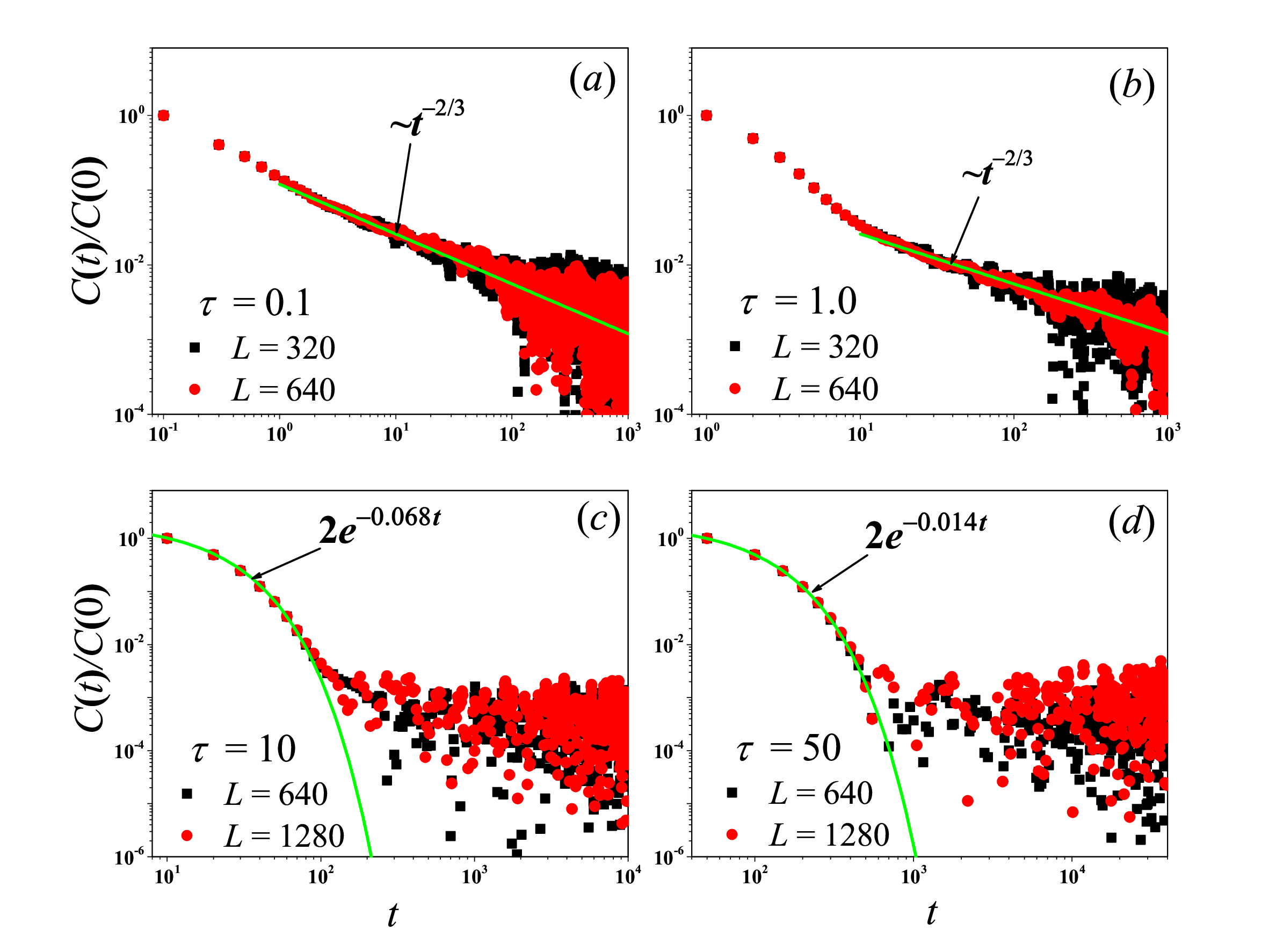}
\caption{(Color online) Correlation functions $C\left(t\right)$ of the total heat current for the 1D fluid system with different $\tau$ values. For reference the green solid line is the best fitting function for the data.}
\label{figct1}
\end{figure}

\begin{figure}
\includegraphics[width=8.5cm]{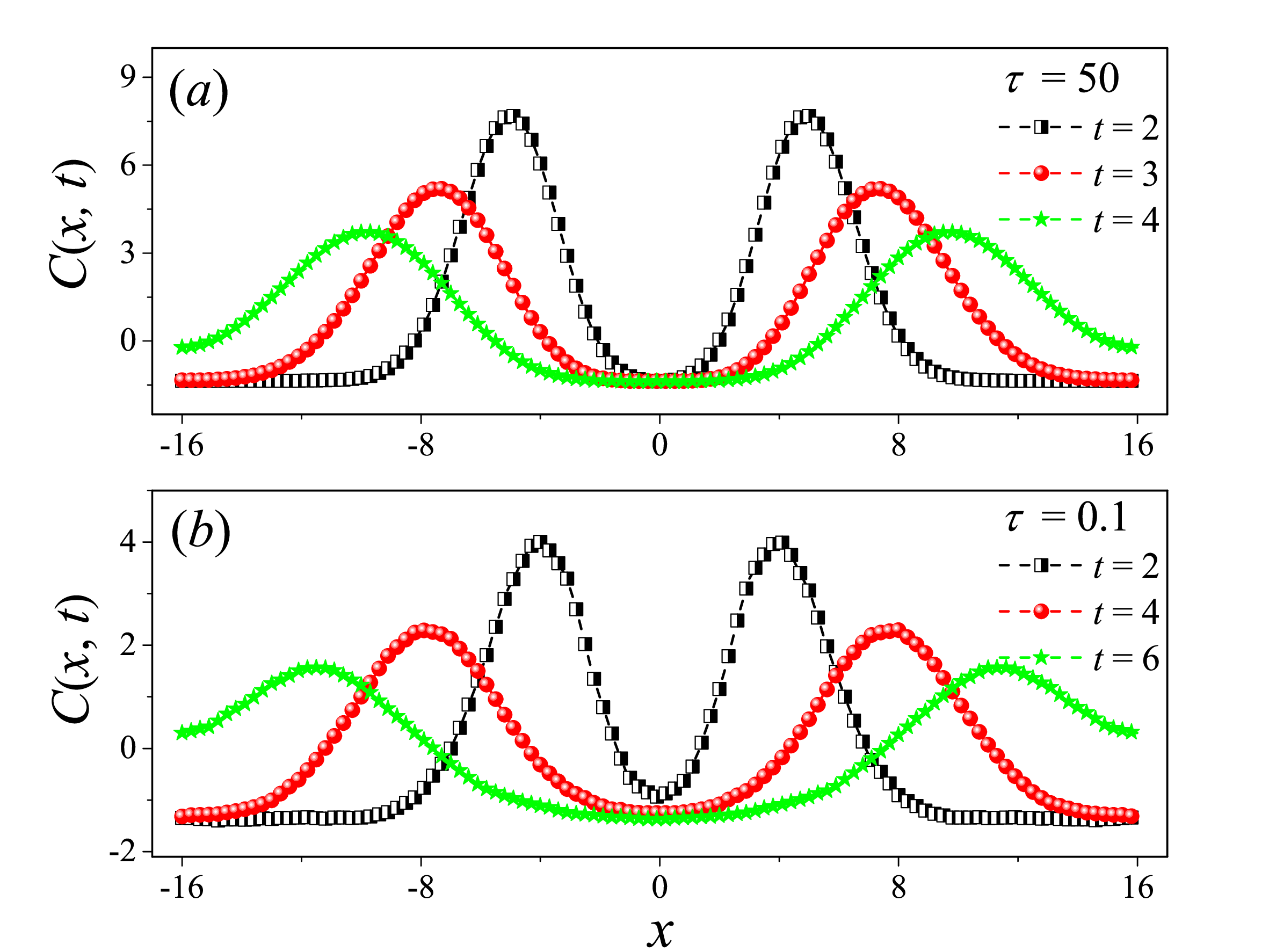}
\caption{ Numerical calculation of the spatiotemporal correlation function $C\left(x,t\right)$ for the 1D fluid system with $\tau=50$ (a) and $\tau=0.1$ (b). Here the system length is set to be $L=32$. One can clearly see the two peaks (the hydrodynamic mode of sound) moving oppositely away from $x=0$ in (a) and (b).}
\label{figCxt1}
\end{figure}

To check the results obtained in the 1D nonequilibrium modeling, we now turn to the comparison with the results obtained by the Green-Kubo formula in the equilibrium modeling. The results for the current time-correlation functions $C(t)$ with different $\tau$ values are presented in Fig.~\ref{figct1}. It can be seen from Fig.~\ref{figct1} (a) and (b) that for $\tau=0.1$ and $\tau=1$, the correlation function eventually attains a power-law decay $C(t)\sim t^{\gamma}$ with $\gamma=-2/3$, fully compatible with the theoretical prediction of the 1D case~\cite{2012Beijeren,2014Spohn}. Substituting it in~Eq.~(\ref{EqGKKL}), and cutting off the integration
as explained above, one obtains the superdiffusive
scaling $\kappa_{\mathrm{GK}}\sim L^{1/3}$, in agreement with our nonequilibrium modeling.

However, it is clear in Fig.~\ref{figct1} (c) and (d) that for $\tau=10$ and $\tau=50$, $C(t)$ undergoes a rapid decay at short times, and eventually, it begins to oscillate around zero (the negative values of $C(t)$ are not shown in this log-log scale.). Fitting function with the green solid line exhibits an exponential decay. To compare $\kappa_{\mathrm{GK}}$ and $\kappa$ more accurately, we compute the sound speed $v_{s}$ of the system with the help of the spatiotemporal correlation function $C\left(x,t\right)$ of local heat currents defined in~Eq.~(\ref{EqCxt}). In Fig.~\ref{figCxt1}, we present $C\left(x,t\right)$ for the system size $L=32$. The two peaks representing the sound mode can be clearly identified in Fig.~\ref{figCxt1} (a) and (b). Their moving speed $v_{s}$ is measured to be $v_s\simeq2.40$ for $\tau=50$ and $v_s\simeq1.875$ for $\tau=0.1$. In Fig.~\ref{figkL1}, the horizontal line for $\tau=50$ is obtained truncating the integral in~Eq.~(\ref{EqGKKL}) upto $L/v_s$
with $v_s=2.40$.  It can be seen that $\kappa_{\mathrm{GK}}(L)$ agrees with $\kappa$. Thus, the equilibrium simulations are fully consistent with the nonequilibrium simulations.

\section{2D fluid}
\label{sec:2D}
\subsection{Nonequilibrium results}
\begin{figure}
\includegraphics[width=9cm]{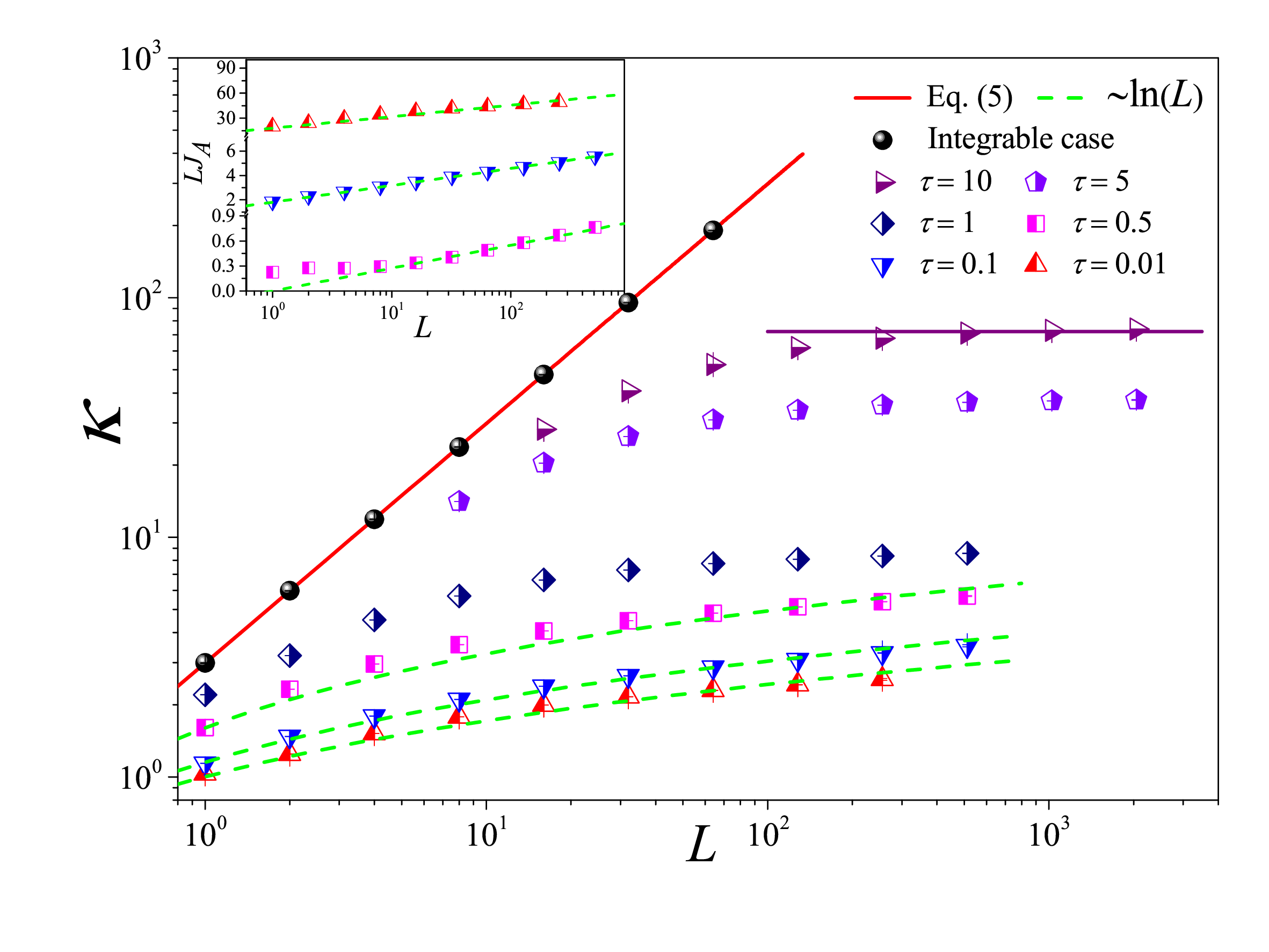}
\caption{(Color online) The thermal conductivity $\kappa$ as a function of the system length $L$ for the 2D square fluid system with different $\tau$ values. The symbols are for the numerical results, and for reference the green dashed line indicates the divergence with $L$ as $\sim\ln(L)$. For $\tau=10$, the horizontal line denotes the saturation value of $\kappa_{\mathrm{GK}}$ obtained by Eq.~(\ref{EqGKKL}), where the integration is up a time $L/v_s$ with $v_s$ measured in Fig.~\ref{figCxt2}. The inset: the log-linear scale is plotted to appreciate the relationship between the product of the anomalous flux  $J_A$ and $L$ and $L$. Here we set $W=L$.}
\label{figkL2}
\end{figure}

\begin{figure}
\includegraphics[width=9cm]{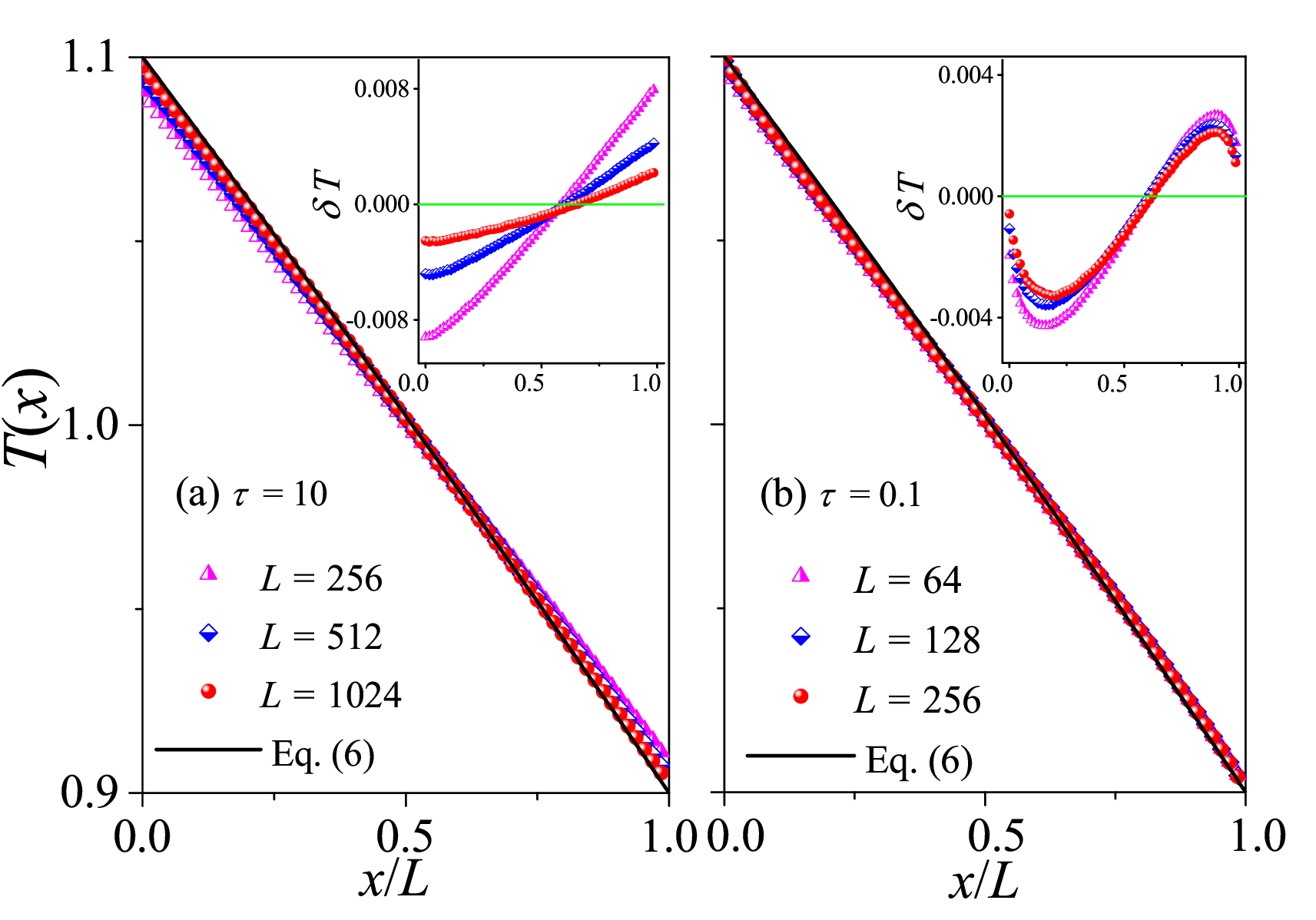}
\caption{(Color online) Plot of temperature profiles $T(x)$ for the 2D square fluid system with different $L$ values. Here our numerical results are compared with the analytical Eq.~(\ref{EqTx}). In (a) and (b) we fix $\tau=10$ and $\tau=0.1$, respectively. Inset: Plot of the differences $\delta T$ between the data and the black line, and the green line at $\delta T = 0$ are for reference.}
\label{figTx2}
\end{figure}

As for the 1D case, let us first reports the dependence of $\kappa$ on $L$ for different $\tau$ values, see Fig.~\ref{figkL2}.
To provide a numerical check, we show that the noninteracting (integrable)
expression Eq.~(\ref{EqKL}) (red solid line) matches very accurately the simulation
data (black circles).
For the interacting systems
it can be seen in Fig.~\ref{figkL2} that for the weak interactions ($\tau=10, 5$), $\kappa$ tends to saturate and becomes constant as $L$ is increased. This indicates that the normal heat conduction behavior for the nearly integrable 2D fluid system is also dominated by kinetic effect. However, as $\tau$ decreases, $\kappa$ is no longer constant but diverges with $L$. In particular, for $\tau=0.1$ and $\tau=0.01$, $\kappa$ eventually approaches the scaling $\kappa\sim \ln(L)$ like that predicted in 2D momentum-conserving systems~\cite{2006Basile,2016Lepri}.
Therefore the scenario is similar to the 1D case and can be described as a crossover from
kinetic to hydrodynamic regimes.

In order to demonstrate the crossover behavior,
we assume  that the flux can be decomposed
as the sum into normal and  anomalous contributions \citep{2020Lepri}
\begin{equation}
J=J_N+J_A,
\label{JA}
\end{equation}
From kinetic theory we expect that
$J_N(L) = \mathfrak{a}/(\mathfrak{b}+L)$ so we can use the conductivity
data for large $\tau$  to estimate $J_N$ and thus
deduce $J_A=J-J_N$.  In the inset of Fig.~\ref{figkL2} we show that for  the strong interactions ($\tau<0.5$), indeed $J_A$ is proportional to $\log (L)/L$. To our knowledge, this provides one of most convincing
numerical evidence of the logarithmic divergence of the conductivity in 2D.

The difference between normal and abnormal heat conduction behaviors can be further verified by $T(x)$ as in the 1D case. For systems with normal heat conduction $T(x)$  is predicted by Eq.~(\ref{EqTx}). In Fig.~\ref{figTx2}(a) this prediction is compared with our simulation results for $\tau=10$. As expected, there is a good agreement between the results of our numerical simulations and Eq.~(\ref{EqTx}). However, for $\tau=10$, $T(x)$ is concave downwards in the left part of the system and concave upwards in the right part of the system. This again conforms to the temperature distribution characteristics of abnormal heat conduction. \par

\subsection{Equilibrium results}

\begin{figure}
\includegraphics[width=9cm]{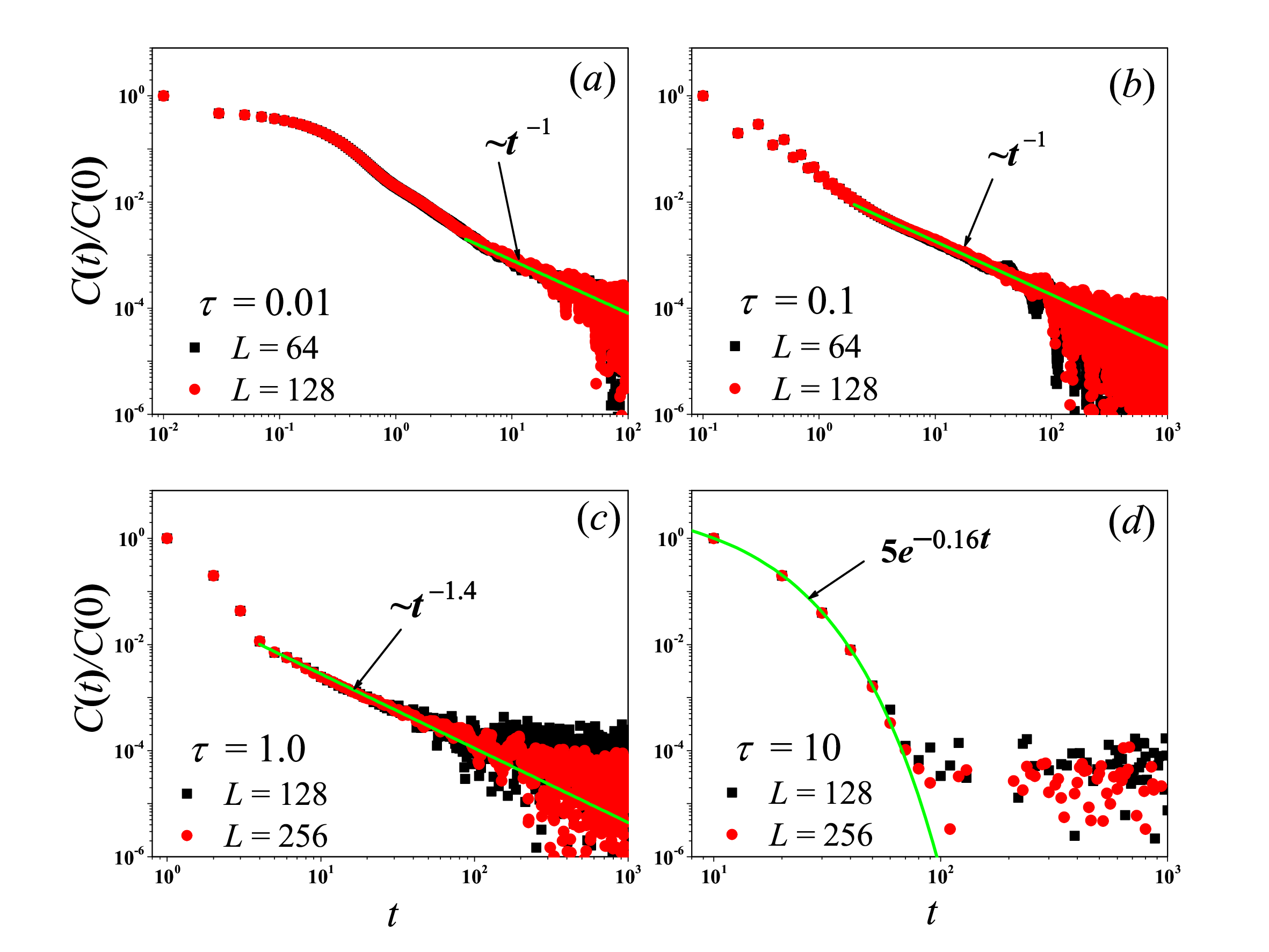}
\caption{(Color online) Correlation functions $C\left(t\right)$ of the total heat current for the 2D square fluid system with different $\tau$ values. For reference the green solid line is the best fitting function for the data. Here we set $W=L$.}
\label{figct2}
\end{figure}

\begin{figure}
\includegraphics[width=8.5cm]{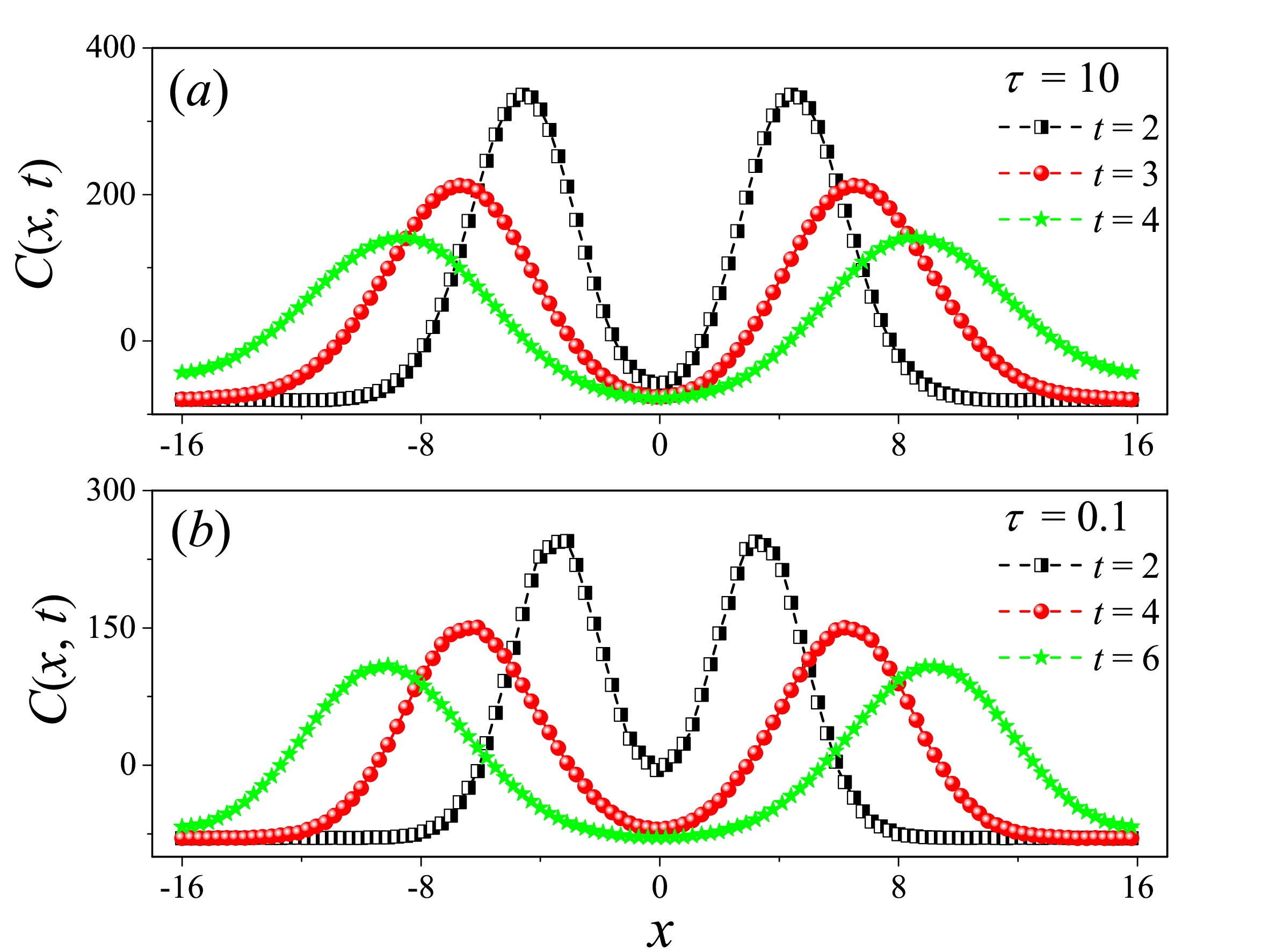}
\caption{ Numerical calculation of the spatiotemporal correlation function $C\left(x,t\right)$ for the 2D square fluid system with $\tau=10$ (a) and $\tau=0.1$ (b). Here we set $W=L=32$. One can clearly see the two peaks (the hydrodynamic mode of sound) moving oppositely away from $x=0$ in (a) and (b).}
\label{figCxt2}
\end{figure}

To check the results obtained above, we now turn to the comparison with the results obtained by the Green-Kubo formula in the equilibrium modeling. The results for $C(t)$ with different $\tau$ values are presented in Fig.~\ref{figct2}. It can be seen from Fig.~\ref{figct2} (a) and (b) that for $\tau=0.01$ and $\tau=0.1$, the correlation function eventually attains a power-law decay $C(t)\sim t^{\gamma}$ with $\gamma=-1$, fully compatible with the theoretical prediction of the 2D case~\cite{2006Basile,2016Lepri}. Taking it in~Eq.~(\ref{EqGKKL}), one will obtain the superdiffusive heat transport $\kappa_{\mathrm{GK}}\sim \ln(L)$, in agreement with nonequilibrium data.

However, it is clear in Fig.~\ref{figct2} that as $\tau$ further increases from $\tau=0.1$ to $\tau=10$, $C(t)$ will change from power-law decay to exponential decay. This means that as $\tau$ increases, the kinetic effects will play a dominant role, and thus heat conduction will change from anomalous behavior to normal behavior, as observed in Fig.~\ref{figkL2}. \par

To compare $\kappa_{\mathrm{GK}}$ and $\kappa$ more accurately, we compute the sound speed $v_{s}$ of the system as performed in 1D case. In Fig.~\ref{figCxt2}, we also present $C\left(x,t\right)$ for the system size $L=32$. The two peaks representing the sound mode can be clearly identified in Fig.~\ref{figCxt2} (a) and (b). Their moving speed $v_{s}$ is measured to be $v_s\simeq2.05$ for $\tau=10$ and $v_s\simeq1.50$ for $\tau=0.1$. In Fig.~\ref{figkL2}, the horizontal line for $\tau=10$ is obtained truncating the integral in~Eq.~(\ref{EqGKKL}) upto $L/v_s$ with $v_s=2.05$.  It can be seen that $\kappa_{\mathrm{GK}}(L)$ agrees with $\kappa$. Again, the equilibrium simulations are fully consistent with the nonequilibrium simulations.
\section{3D fluid}
\label{sec:3D}

\subsection{Nonequilibrium results}

\begin{figure}
\includegraphics[width=9cm]{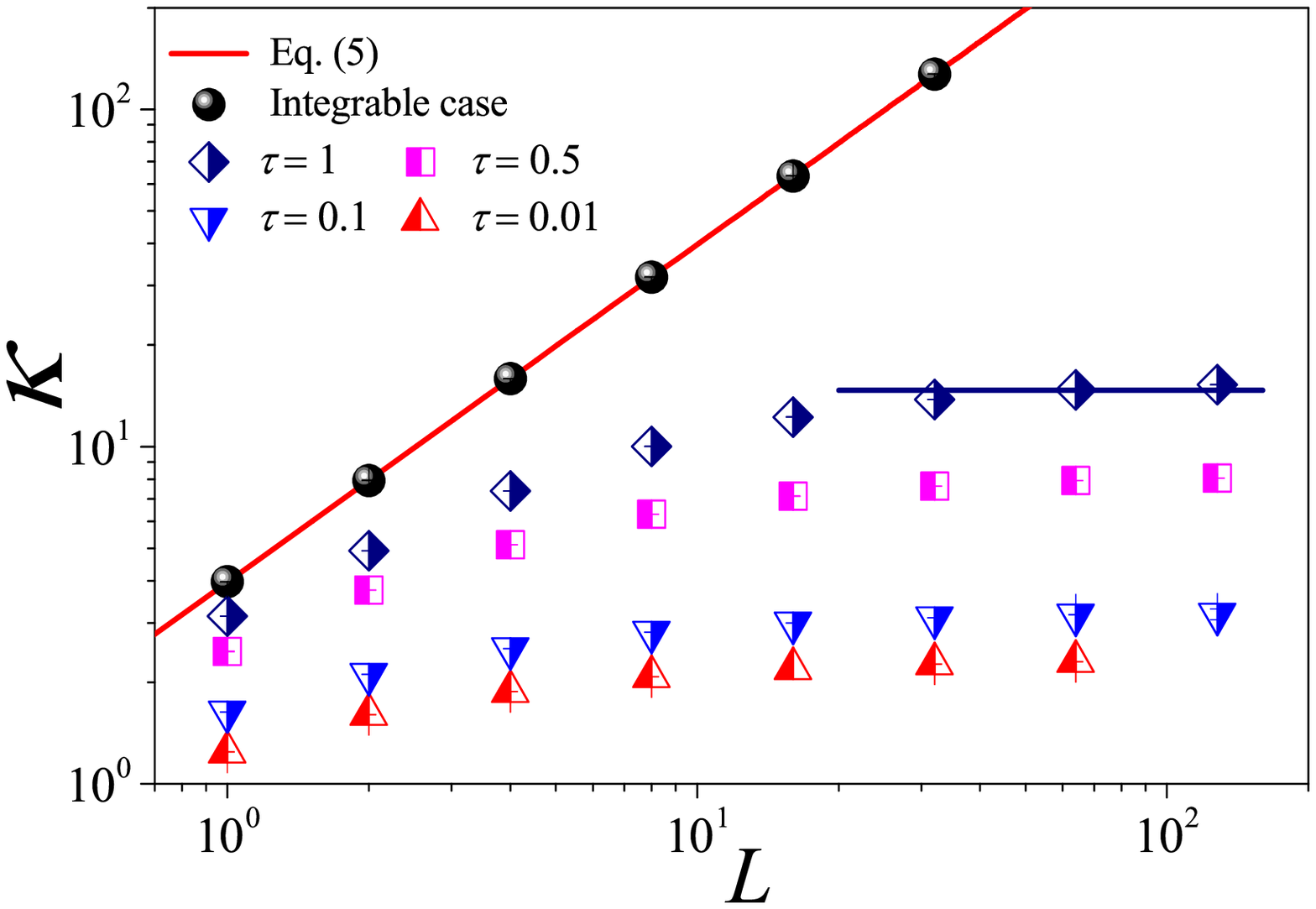}
\caption{(Color online) The thermal conductivity $\kappa$ as a function of the system length $L$ for the 3D cubic fluid system with different $\tau$ values. Here we set $W=H=L$.}
\label{figkL3}
\end{figure}

To complete the study of dimensionality effects, we also performed a series of simulations
for the 3D case. The results in Fig.~\ref{figkL3} demonstrate that even here the
formula Eq.~(\ref{EqKL}) accounts very accurately for the non-interacting case
(compare the red solid line with black circles).
The data for the interacting case (symbols in Fig.~\ref{figkL3}) confirms that, even for
the smallest $\tau$ considered the conductivity converges to a finite value.
As expected, Fourier's law holds for 3D fluid system with the momentum conservation.

\begin{figure}
\includegraphics[width=9cm]{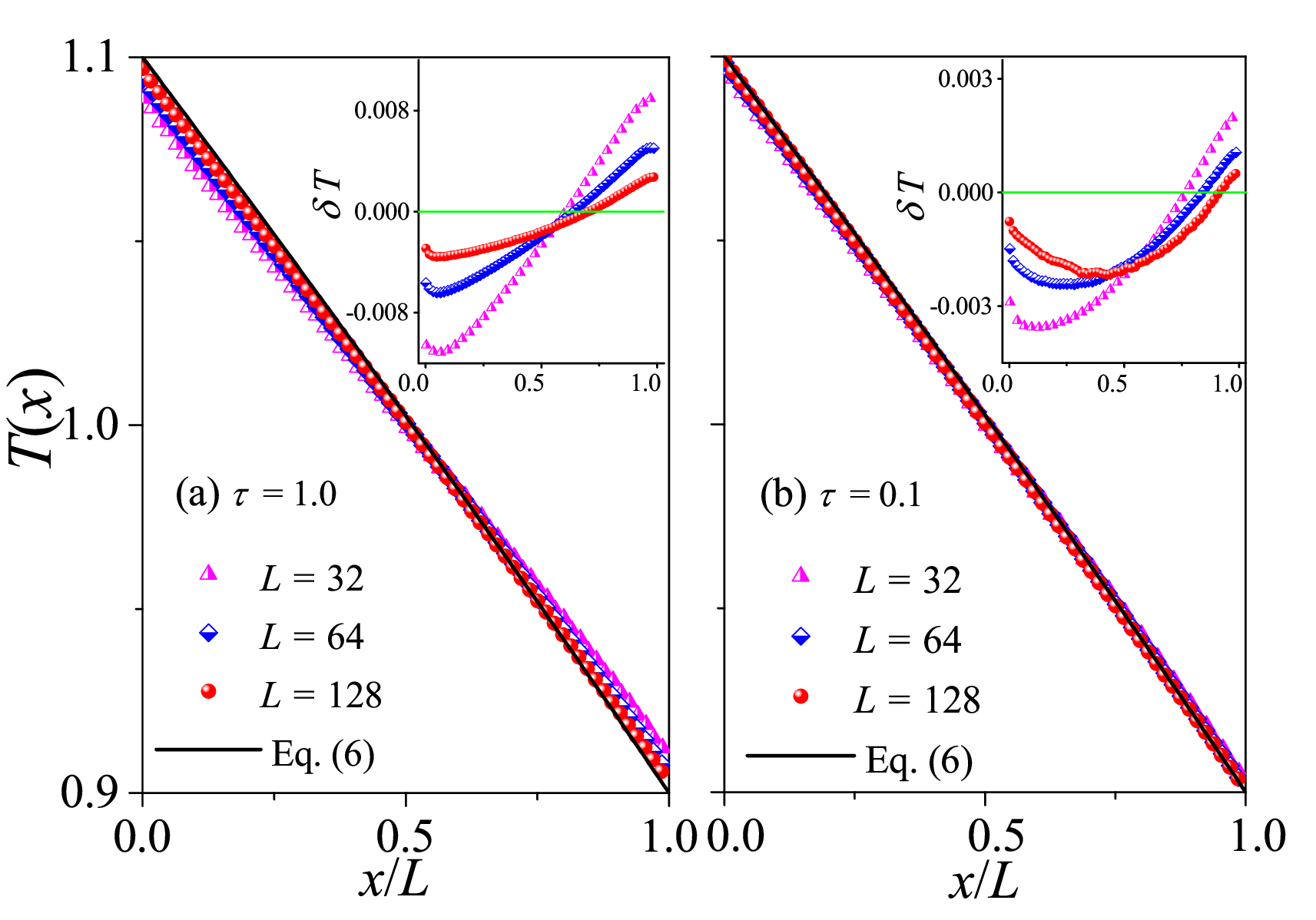}
\caption{(Color online) Plot of temperature profiles $T(x)$ for the 3D cubic fluid system with different $L$ values. Here our numerical results are compared with the analytical Eq.~(\ref{EqTx}). In (a) and (b) we fix $\tau=1.0$ and $\tau=0.1$, respectively. Inset: Plot of the differences $\delta T$ between the data and the black line, and the green line at $\delta T = 0$ are for reference.}
\label{figTx3}
\end{figure}

The normal heat conduction behavior can be further verified by $T(x)$. For systems with normal heat conduction $T(x)$  is predicted by Eq.~(\ref{EqTx}). In Fig.~\ref{figTx3} this prediction is compared with our simulation results for $\tau=1$ and $\tau=0.1$. As expected, there is a good agreement between the results of our numerical simulations and Eq.~(\ref{EqTx}), conforming to the temperature distribution characteristics of normal heat diffusion.  \par

\subsection{Equilibrium results}
\begin{figure}
\includegraphics[width=10cm]{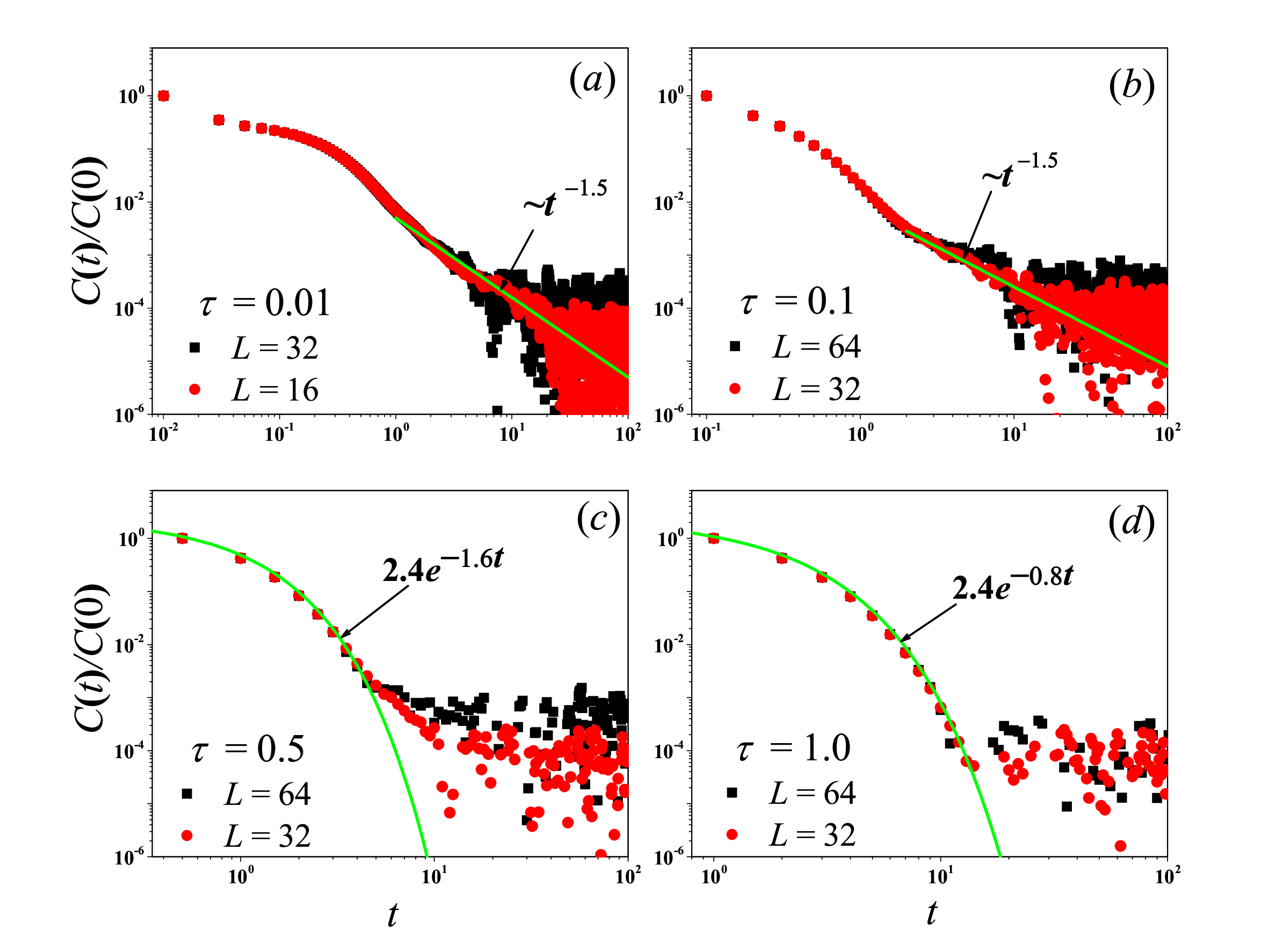}
\caption{(Color online) Correlation functions $C\left(t\right)$ of the total heat current for the 3D cubic fluid system with different $\tau$ values. For reference the green solid line is the best fitting function for the data. Here we set $W=H=L$.}
\label{figct3}
\end{figure}

\begin{figure}
\includegraphics[width=8.5cm]{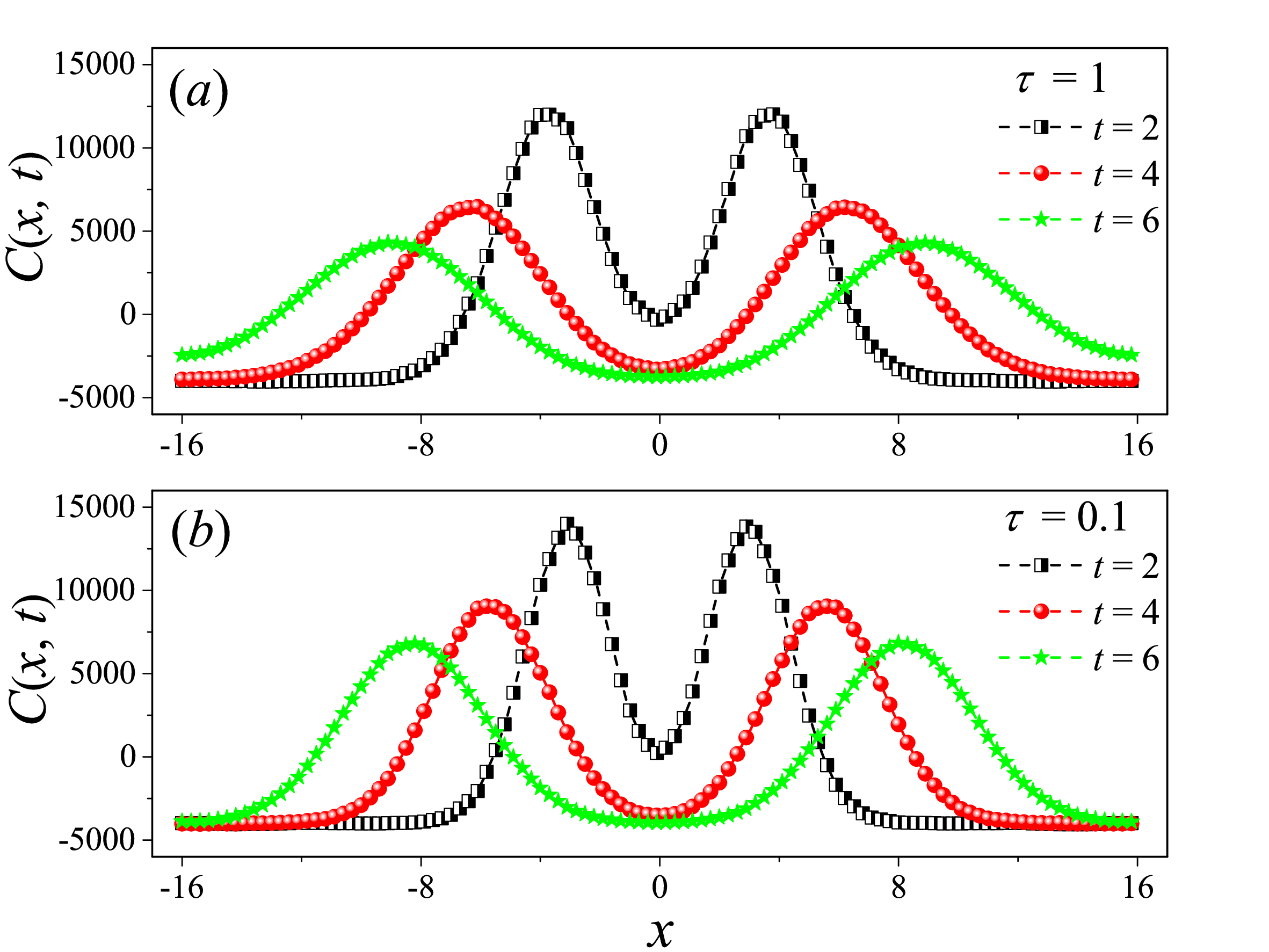}
\caption{ Numerical calculation of the spatiotemporal correlation function $C\left(x,t\right)$ for the 3D cubic fluid system with $\tau=1$ (a) and $\tau=0.1$ (b). Here we set $W=H=L=32$. One can clearly see the two peaks (the hydrodynamic mode of sound) moving oppositely away from $x=0$ in (a) and (b).}
\label{figCxt3}
\end{figure}

To check what obtained in the 3D nonequilibrium modeling, we now turn to the comparison with the results obtained by the Green-Kubo formula in the equilibrium modeling. The results for $C(t)$ with different $\tau$ values are presented in Fig.~\ref{figct3}. It can be seen from Fig.~\ref{figct3} (a) and (b) that for $\tau=0.01$ and $\tau=0.1$, the correlation function eventually attains a power-law decay $C(t)\sim t^{\gamma}$ with $\gamma=-1.5$, fully compatible with the theoretical prediction of the 3D case~\cite{2006Basile,2016Lepri}.\par

However, it is clear in Fig.~\ref{figct3} that as $\tau$ further increases from $\tau=0.01$ to $\tau=1.0$, $C(t)$ will change from power-law decay to exponential decay. This means that as $\tau$ increases, the normal heat conduction behavior  observed in Fig.~\ref{figkL3} will change from being dominated by the hydrodynamic effect to being dominated by the kinetic effect. \par

To compare $\kappa_{\mathrm{GK}}$ and $\kappa$ more accurately, we compute the sound speed $v_{s}$ of the 3D system. In Fig.~\ref{figCxt3}, we also present $C\left(x,t\right)$ for the system size $L=32$. The two peaks representing the sound mode can be clearly identified in Fig.~\ref{figCxt3} (a) and (b). Their moving speed $v_{s}$ is measured to be $v_s\simeq1.30$ for $\tau=1$ and $v_s\simeq1.275$ for $\tau=0.1$. In Fig.~\ref{figkL3}, the horizontal line for $\tau=1$ is obtained truncating the integral in~Eq.~(\ref{EqGKKL}) upto $L/v_s$ with $v_s=1.30$.  It can be seen that $\kappa_{\mathrm{GK}}(L)$ agrees with $\kappa$. Once again, the equilibrium simulations are fully consistent with the nonequilibrium simulations.

\section{Dimensional crossovers}
\label{sec:cross}

\begin{figure}
\includegraphics[width=9cm]{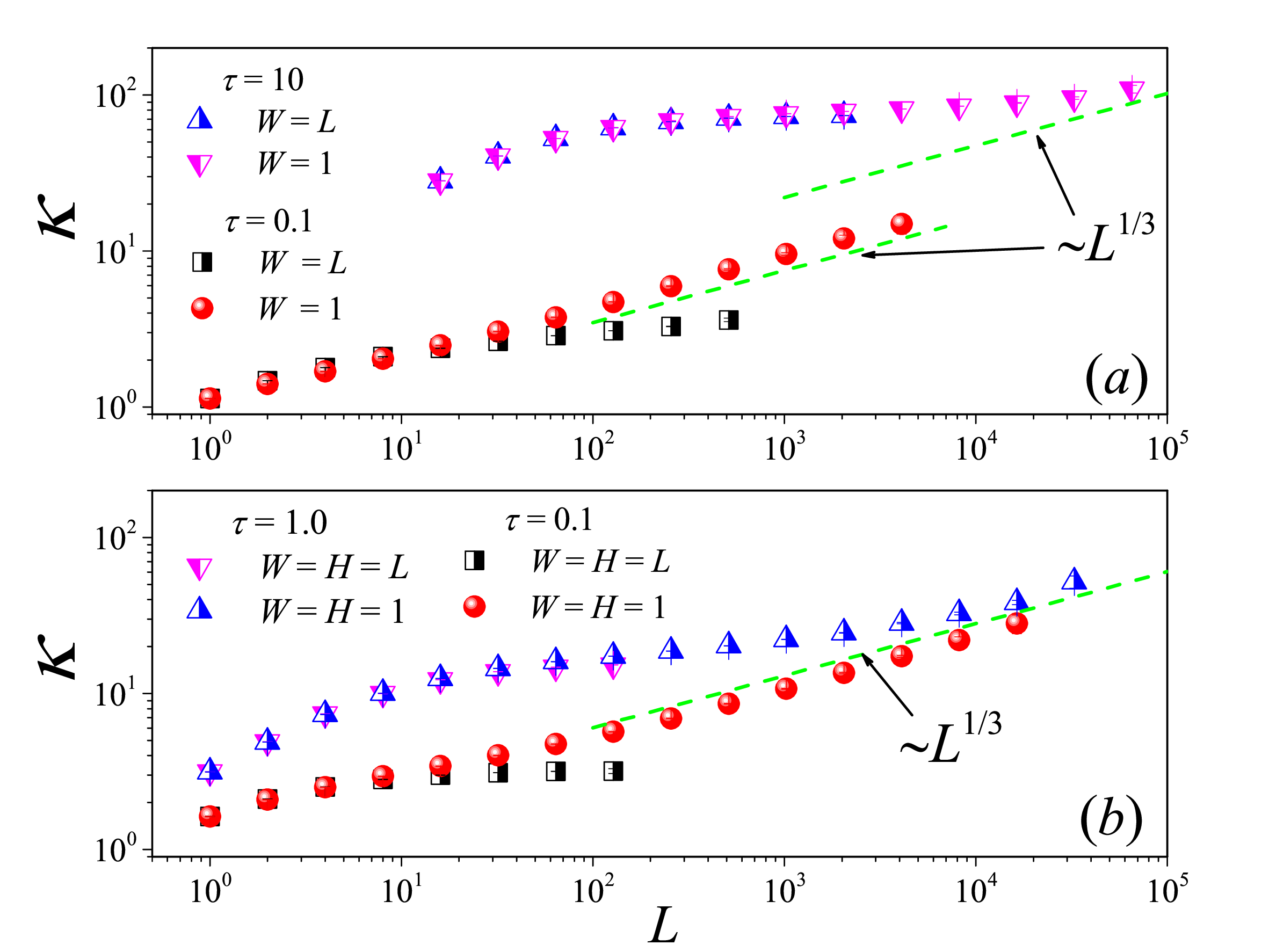}
\caption{(Color online) The dimensional crossover behavior of heat conduction for the 2D (a) and 3D (b) fluid system with different $\tau$ values.}
\label{figkL4}
\end{figure}

Dimensional-crossover   is a relevant topic for thermal transport in low-dimensional
materials~\cite{2018Gu}. Indeed, in 2014 it has been experimentally observed in the suspended single-layer graphene~\cite{2014Xu}. In this experimental setup, the width of the samples is
kept fixed and the thermal conductivity changes upon increasing their length
is measured. As the length increases, it is natural to expect that a dimensional-crossover behavior from two dimensions to quasi-one dimension will occur. These research results have greatly enriched our understanding of heat conduction in lattice systems.\par

We show that the MPC approach can be used successfully to investigate this issue, considering
2D and 3D mesoscopic fluid models with fixed transverse sizes ($W=1$ in 2D and $W=H=1$ in 3D) and study how $\kappa$ changes with $L$. It can be seen from Fig.~\ref{figkL4} (a) that in 2D fluid models, upon increasing the aspect ratio of the system, for both $\tau=10$ and $\tau=0.1$, $\kappa$  eventually follows the 1D divergence law $\kappa\sim L^{1/3}$  as $L$ increases. This means that both in the case where dominated by the kinetic  ($\tau=10$) or hydrodynamic effect ($\tau=0.1$), there exists dimensional-crossover above a given aspect ratio. As is shown in Fig.~\ref{figkL4} (b), in 3D fluid models, there is also a similar phenomenology. Altogether, these results confirm that
the theories developed for the strictly 1D case effectively extend also
quasi-1D, provided that the transverse extent of the sample is small enough. \par

\begin{figure}
\includegraphics[width=9cm]{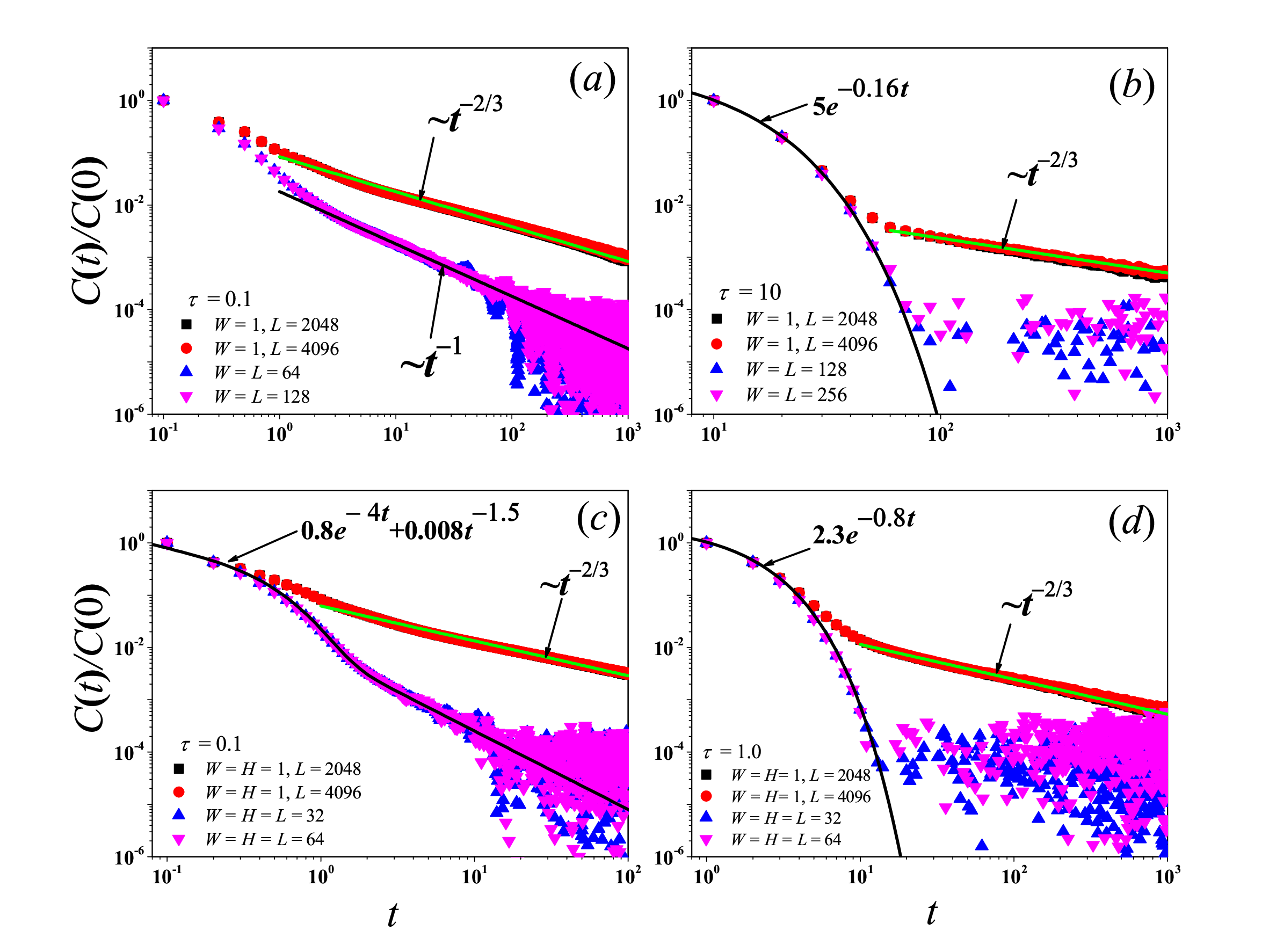}
\caption{(Color online) Correlation functions $C\left(t\right)$ of the total heat current for the 2D and 3D fluid system with different $\tau$ values. For reference, the green and black solid lines are the best fitting functions for the data.}
\label{figct4}
\end{figure}

To further support the dimensional-crossover behavior of heat conduction observed above, we perform the equilibrium simulations of  $C(t)$ in 2D and 3D fluid models. The results for $C(t)$ with different $\tau$ values are presented in Fig.~\ref{figct4}. We can see from Fig.~\ref{figct4} (a) and (b) that in 2D fluid models, under the condition of increasing the aspect ratio of the system, for heat conduction dominated by the hydrodynamic effect ($\tau=0.1$) or dominated by the kinetic effect ($\tau=10$), $C(t)$ will eventually change to a power-law decay $C(t)\sim t^{-2/3}$. As is shown in Fig.~\ref{figct4} (c) and (d), in 3D fluid models, there is also a similar phenomenon that $C(t)$ will eventually change to a power-law divergence for the hydrodynamic effect ($\tau=0.1$) and the  kinetic effect ($\tau=1$). \par

\section{Heat transfer with magnetic field}
\label{sec:magnetic}

Another issue that can be studied through the MPC dynamics concerns the influence of a magnetic field on transport~\cite{2017Saito,2018Saito,2018Tamaki,2022Bhat}. It is generally believed that heat conduction behavior is normal in low-dimensional systems where momentum is not conserved~\cite{2008Dhar,2016Lepri}. However, there is a counterexample in a low-dimensional system with magnetic field. Specifically, heat transport via the one-dimensional charged particle systems with transverse motions is studied in~\cite{2017Saito}, where researchers studied two cases: case (I) with uniform charge and case (II) with alternate charge. An intriguing finding of this study is that in both cases involving non-zero magnetic fields, the heat conduction behaviors exhibit anomalies, similar to the case where momentum is conserved under the zero magnetic field condition. Remarkably, the abnormal behavior in case (I) is different from the case without magnetic field, suggesting a novel dynamical universality class. Due to the presence of the magnetic field, the standard momentum conservation in such a system is no longer satisfied but is replaced by the pseudomomentum conservation~\cite{1983Johnson}. Thus, there are two relevant questions: (1) Does the pseudomomentum conservation of a system lead to abnormal heat conduction? (2) Can the abnormal behaviors in both cases also be observed in low-dimensional fluids under the same pseudomomentum conservation?\par
\begin{figure}
\includegraphics[width=9cm]{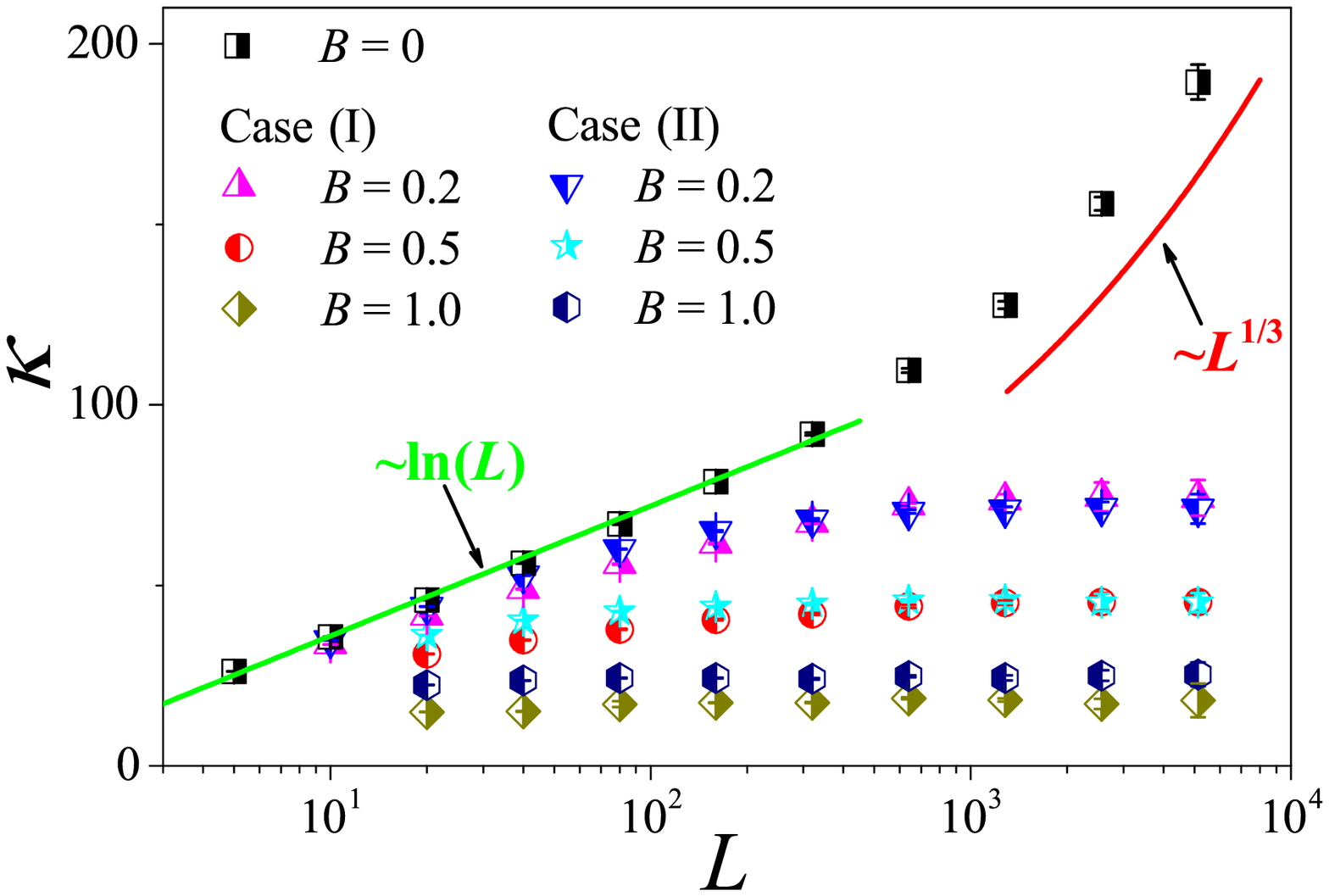}
\caption{The heat conductivity $\kappa$ as a function of the system length $L$ for the 2D fluid system without and with a magnetic field. The symbols are for the numerical results. For reference the green straight line is the best logarithmic fit, $\kappa\sim \ln (L)$ and the red curve line indicates the divergence with $L$ as $\sim L^{1/3}$. Except for  $W=a=\tau=0.1$ and $\rho=22$, other parameters are consistent with those adopted in this paper.}
\label{figm1}
\end{figure}
The above two questions have been well answered only recently in our research~\cite{2025luo}, where it is shown that under the same pseudomomentum conservation, the 2D fluid system with magnetic field can exhibit normal heat conduction behavior. Specifically, we consider a 2D system of charged particles as depicted in Fig.~\ref{model} (b). In this system, a constant magnetic field perpendicular to the plane of motion, $\textbf{\emph{B}}=B\textbf{\emph{k}}$, is imposed. The particles interact via the modified MPC dynamics to maintain the pseudomomentum conservation of the system (see~\cite{2025luo} for details). To compare with the results obtained in~\cite{2017Saito}, we also consider two cases: case (I) with uniform charges $e_i=1$ and case (II) with opposite charges on each half of particles, say $e_i=(-1)^i$. In Fig.~\ref{figm1}, we plot the relation of $\kappa$ vs $L$ for various $B$ obtained by nonequilibrium thermal-wall method. It is shown that for $B=0$, the system with momentum conservation exhibits the crossover from 2D to 1D behavior of the thermal conductivity under the condition of increasing the aspect ratio of the system. However, for $B\neq0$, heat conduction behaviors in both cases with pseudomomentum conservation are normal because as $L$ increases, $\kappa$ approaches a finite value.\par

Obviously, our above results are at variance with the findings in~\cite{2017Saito}.
There, heat conduction in presence of pseudomomentum conservation in two cases are abnormal. This observation, together with our results, thus clarify that pseudomomentum conservation is not related to the normal and anomalous behaviors of heat conduction and provide an example of the difference in heat conduction between fluids and lattices in the presence of the magnetic field condition.

For completeness, we also mention that the MPC scheme can be extended to the case of
charged particles that yields a self-consistent electric field. This situation
is relevant for plasma physics and can be treated by coupling the MPC
dynamics with a Poisson solver (see \citep{dicintio2017multiparticle} and references therein
for details). The effect of the electric field on heat transport can be studied
by this method: simulations reveal that the field does not affect
significantly the hydrodynamics
of the model, at least for not too large amplitude  fluctuations \citep{dicintio2017multiparticle}.

\section{Conclusions}

We have presented a series of numerical simulation demonstrating how the MPC  method
can be effectively employed to study the dimensionality effects on heat transfer
in a simple confined fluid. Nonequilibrium dynamics can be simulated efficiently
with the thermal-wall modeling of external reservoir and the results agree very
well with Green-Kubo linear response.
The data are statistically very accurate and span
over a considerable range of system sizes. The overall theoretical scenario
is confirmed by the data. It should be noticed that most of the publications
in this context refer to lattice systems \citep{2016Lepri}, so our results represent are a relevant
extension to the case where particle are free to diffuse through the simulation
box. This supports the general validity of low-dimensional hydrodynamic theories
and of the tight connection with Kardar-Parisi-Zhang physics in transport problems
\citep{2014Spohn}.

Another relevant finding is that the crossover from diffusive to anomalous regimes,
seen in quasi-integrable chains \citep{2020Lepri}, extends to the somehow simpler
case of fluids. In particular the decomposition of the current Eq.(\ref{JA})
is an effective and simple way to assess the divergence law when interaction
is relatively weak and the accessible range of sizes too limited (as
frequently occurs in practice).

We also illustrated how the important issue of dimensional crossovers and
the effect of an applied magnetic field can be studied relatively easily
via the MPC dynamics. A further extension would be to introduce the
effect of chemical baths, namely to account for the exchange of particles
with environment \cite{benenti2014thermoelectric,2018luo}. This would allow
to study basic features of coupled transport process but also
design and conceive novel possible applications.

\section*{Acknowledgment}
Useful discussions with Weicheng Fu are gratefully acknowledged. We acknowledge support by the National Natural Science Foundation of China (Grants No.12475034, No.12465010, and No.12105049) and the Natural Science Foundation of Fujian Province (Grant No.2023J05100).

\bibliography{paper}

\end{document}